\documentclass[%
 reprint,
 amsmath,amssymb,
 aps,
]{revtex4-1}
\usepackage{color}
\usepackage{graphicx}
\usepackage{dcolumn}
\usepackage{bm}

\begin{document}

\title{Entanglement, Superselection Rules and Supersymmetric \\ Quantum Mechanics}

\author{E. Cattaruzza}
 \email{Enrico.Cattaruzza@gmail.com}
\affiliation{INFN, Sezione di Trieste, Trieste, Italy.}
\author{E. Gozzi}%
 \email{ennio.gozzi@ts.infn.it}
\affiliation{Dipt. di Fisica, Sezione Teorica, Universit\'a di Trieste, \\ Strada Costiera 11, Miramare-Grignano, 34151, Trieste, Italy\\ and \\
INFN, Sezione di Trieste, Trieste, Italy.}%
\author{Carlo Pagani} 
 \email{carlo.pagani@sissa.it}  
\affiliation{SISSA, Via Bonomea 265; Trieste, 34014, Italy.}



\date{\today}

\begin{abstract}
In this paper we show that the energy eigenstates of supersymmetric quantum mechanics (SUSYQM) with non-definite ``fermion" 
number are entangled states. They are "{\it physical states}"  of the model provided that observables with odd number of spin variables are allowed in the theory like it happens in the Jaynes-Cummings model. Those states generalize the so called {\it "spin-spring"} states of the Jaynes-Cummings model which have played an important role in the study of entanglement. 
\begin{description}
\item[PACS numbers] 03.65.-w, 03.65.ud, 11.30.Pb
\end{description}
\end{abstract}

\maketitle


\section{Introduction}
Entanglement is quite a  unique feature of quantum mechanics discovered by E.Scroedinger in 1935~\cite{Schroedinger}. Its 
study has deepened over the last 15 years because of its potential application to quantum information and quantum computation \cite{Nielsen}.
Experimentally entangled states have been realized and built mainly in quantum optics \cite{Haroche}.  In this field a phenomenological model 
which has had a great success has been the Jaynes-Cummings model (JC) \cite{Cummings}. 
This model somehow describes a laser wave interacting with a two-level atom. Its Hamiltonian has become the prototype of 
what are called the ``{\it spin-spring}'' systems. These are made of a spin-system (or two level atomic system) interacting with a spring-system (or equivalently a wave). 
The crucial output are the entangled states which naturally appear in this model. 
\par We thought that a natural generalization of the ``{\it spin-spring}'' systems are what we could call ``{\it spin-potential}'' systems in which a spin-system interacts with a system living in a generic potential and not necessarily a spring-one. A model of this sort is supersymmetric quantum mechanics
~\cite{Witten} (SUSYQM). Its eigenstates of non definite fermion number turns out to be entangled states which are "{\it physical states}" \cite{Wic}\cite{Rom} of the theory provided that, like in the Janes-Cummings model, observables odd in the spin variables are allowed in the theory, otherwise a superselection mechanism\cite{Wic}\cite{Rom} is triggered which cancels  those states from the physical ones.
\section{Supersymmetric Quantum Mechanics}
This model was introduced in 1981 by E. Witten~\cite{Witten} as a playground on which to test the issue of supersymmetric 
breaking in realistic supersymmetric field theories. Soon the model acquired a life  by itself which shed light on several topics 
which ranged from nuclear physics~\cite{Ginocchio}, to stochastic processes~\cite{Parisi,*Cecotti,*Coop,*Goz}, to differential geometry~\cite{Witten2},  to atomic physics ~\cite{Kostelecky}, to non-perturbative methods~\cite{Alan,*ennD}, and to integrability~\cite{Scg,*Infe,*Rus,*Grou}. In this last topic the work again started with Schroedinger and it went under the name of "factorization method" and not supersymmetry; more references about this interesting topic, which evolved into the so called "shape invariance method", can be found in \cite{Asim}.
 
Nice reviews have been written on Susy Q.M.~\cite{Cooper} and even books~\cite{Junker,Asim}. By consulting them, the reader can get an idea of the many applications of SUSYQM.
\par The 1-dim. Hamiltonian for this system is:
\begin{equation}
\widehat H=\hat p^2+W^2(\hat q)-\left[\hat \psi^{\dag},\hat \psi\right]\,W^{\prime}(\hat q),
\label{(4-1)}
\end{equation}
where $W(q)$ is an arbitrary function of $q$ called super potential and $\hat \psi^{\dag},\hat \psi$ are Grassmannian variables.
 We have neglected the factor $1/2$ in front of $\hat p^2$ in order to simplify things and we have put $\hbar =1$. The commutation relations are:
 \begin{equation}
       \left[\hat p,\hat q\right]_{-}=-1,\,\,\, \left[\hat \psi^{\dag},\hat \psi\right]_{+}=1,\,\,\, \hat \psi^{\dag^2}=\hat \psi^2=0,
\label{(4-2)}
\end{equation}
 where $[\,\,,\,\,]_{\mp}$ indicates respectively the commutator and the anticommutator. The Hamiltonian (\ref{(4-1)}) describes two interacting systems: a ``spin  one-half''
 described by the Grassmannian variables $\hat \psi^{\dag},\hat \psi$ and a ``bosonic'' one described by the $\hat q$ variable. These two ``systems''  are interacting with each other via the term $\left[\hat \psi^{\dag},\hat \psi\right]\,W^{\prime}(\hat q)$ present in (\ref{(4-1)}). The full Hamiltonian has a symmetry known as a $N=2$ supersymmetry where the conserved charges are:
\begin{eqnarray} 
\hat Q&\equiv& \left(\hat p+i\,W(\hat q)\right)\hat \psi^{\dag}\nonumber\\
\hat Q^{\dag}&\equiv& \left(\hat p-i\,W(\hat q)\right)\hat \psi.
 \label{(5-1)} 
\end{eqnarray}
This symmetry is called supersymmetry because the two charges in  (\ref{(5-1)}) close on the Hamiltonian:
\begin{equation}
\left[\hat Q,\hat Q^{\dag}\right]_{+}=\widehat H.
\label{(5-2)}
\end{equation}
These charges ``rotate'' a ``bosonic'' degree of freedom ($\hat q$) into a "fermionic"  one ($\hat \psi,\hat \psi^{\dag}$). In Susy Q.M. another conserved quantity is the ``fermion'' number $(-1)^{\mathbb F}$ defined  \cite{Witten} as: 
\begin{equation}
(-1)^{\mathbb F}\equiv 1-2\,\hat \psi\hat \psi^{\dag}.
\label{(5-3)}
\end{equation}
Let us now move to a more physical realization of the ``spin'' degrees of freedom. The commutation relation for the $\hat \psi,\hat\psi^{\dag}$ appearing in  (\ref{(4-2)}) can be realized via the following matrix representation for the $\hat \psi$ and $\hat \psi^{\dag}$:
\begin{equation}
\hat\psi=\left(\begin{matrix}
0&1\\
0&0
\end{matrix}
\right),\quad 
\hat\psi^{\dag}=\left(\begin{matrix}
0&0\\
1&0
\end{matrix}
\right).
\label{(6-1)}
\end{equation}
Inserting this expression into the Hamiltonian  (\ref{(4-1)}), we get: 
\begin{eqnarray}
\widehat H&=&\hat p^2+W^2(\hat q)+\sigma_3\,W^{\prime}(\hat q)\nonumber\\
&=&\left(\begin{matrix}
\widehat H_+&0\\
0&\widehat H_-
\end{matrix}
\right)
\label{(6-2)}
\end{eqnarray}
where  $\sigma_{3}$   is the third Pauli matrix and 
\begin{eqnarray*}
\widehat H_+&=&\hat p^2+W^2(\hat q)+W^{\prime}(\hat q)\\
\widehat H_-&=&\hat p^2+W^2(\hat q)-W^{\prime}(\hat q). 
\end{eqnarray*}
We can conclude that, via the  matrix representation  (\ref{(6-1)}) for the $\hat \psi$ and $\hat \psi^{\dag}$,
the $\widehat H$ becomes  a $2\times 2$ matrix of Hamiltonians  like the one appearing in the Pauli 
equation and in the Jaynes-Cummings model~\cite{Cummings}.  For the interested reader we should say that SUSYQM models in higher dimensions have been studied in  \cite{Andrianov} while supersymmetric  models with a much richer structure that the ones we have presented in this section have been explored in ref. \cite {13bis}.
\section{Eigenstates and Entanglement}
Let us now turn to SUSYQM models in 1 dimension.
Suppose we are able to exactly diagonalize the Hamiltonian  (\ref{(6-2)}) and that its spectrum is discrete.
Indicating with $\varphi_n^{+},\,\varphi_n^{-}$ the eigenfunctions associated to $\widehat H_+$ and $\widehat H_-$
we can write:
\begin{equation}
\left(\begin{matrix}
\widehat H_+&0\\
0&\widehat H_-
\end{matrix}
\right)
\left(\begin{matrix}
\varphi_n^{+} \\
\varphi_n^{-}
\end{matrix}
\right)=E_n\left(\begin{matrix}
\varphi_n^{+} \\
\varphi_n^{-}
\end{matrix}
\right).
\label{(7-1)}
\end{equation}
This is equivalent to
\begin{equation}
\left\{
\begin{matrix}
\widehat H_+\,\varphi_n^{+}&=&E_n\,\varphi_n^{+}\\
\widehat H_-\,\varphi_n^{+}&=&E_n\,\varphi_n^{-}.
\end{matrix}
\right.
\label{(7-2)}
\end{equation}
These last relations  (\ref{(7-2)}) implies that $\widehat H_+$ and $\widehat H_-$ are isospectral Hamiltonian which means that they have the same spectrum. 
This is true for all levels except for the $E_0=0$ level which is present in the case of Susy not-spontaneosly broken  \cite{Witten}. In this case the $E_0=0$ eigenstate and eigenvalue equation is:
 \begin{equation}
\left(\begin{matrix}
\widehat H_+&0\\
0&\widehat H_-
\end{matrix}
\right)
\left(\begin{matrix}
0 \\
\varphi_0^{-}
\end{matrix}
\right)=0.
\label{(8-1)}
\end{equation}
If, on the contrary, Susy is spontaneously broken \cite{Witten}, then there is no normalizable eigenstate at $E_0=0$. They are all at $E>0$. For more 
details about this see ref. \cite{Witten} and \cite{Junker}. The pictures of the $\widehat H_+$ and $\widehat H_-$ potentials are showed in the two figures below.
\begin{figure}[h]
\includegraphics[scale=0.9]{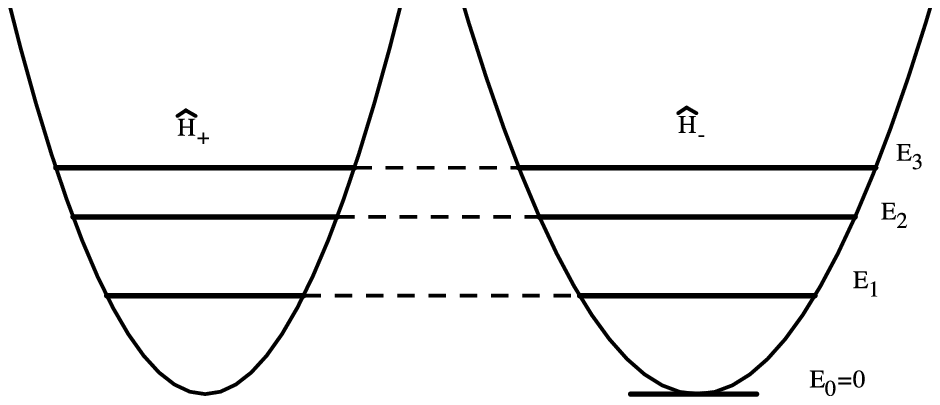}
\caption{\label{Fig1} Susy not spontaneously broken.}
\includegraphics[scale=0.9]{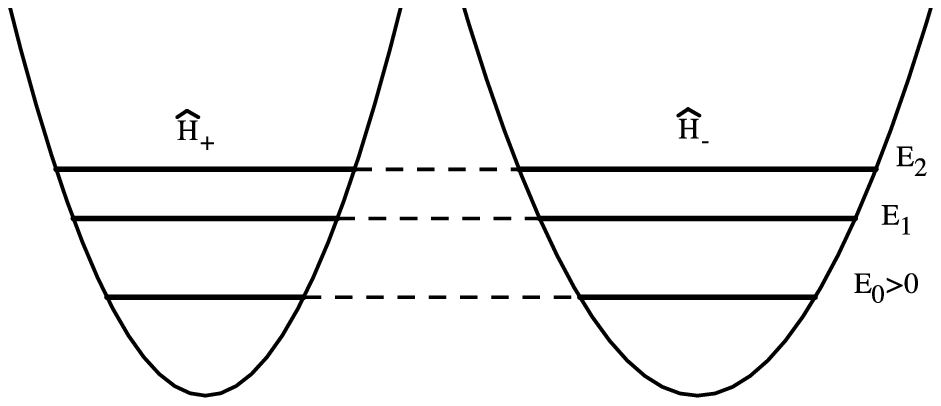}
\caption{\label{Fig2} Susy  spontaneously broken.}
\end{figure}
Let  us now return to the eigenstates  (\ref{(7-1)}) and let us stick to the case $E_n>0$.
It is easy to realize that there are other states, besides those in $(\ref{(7-1)})$, which are at energyy $E_n$.
They are:  
 \begin{equation}
\left(\begin{matrix}
\widehat H_+&0\\
0&\widehat H_-
\end{matrix}
\right)
\left(\begin{matrix}
\varphi_n^{+} \\
0
\end{matrix}
\right)=E_n\left(\begin{matrix}
\varphi_n^{+} \\
0
\end{matrix}
\right).
\label{(9-1)}
\end{equation}
and 
 \begin{equation}
\left(\begin{matrix}
\widehat H_+&0\\
0&\widehat H_-
\end{matrix}
\right)
\left(\begin{matrix}
0 \\
\varphi_n^{-}
\end{matrix}
\right)=E_n\left(\begin{matrix}
0 \\
\varphi_n^{-}
\end{matrix}
\right).
\label{(9-2)}
\end{equation}
This degeneracy is due to supersymmetry. In fact , using the concentions of ref. \cite{ingo}, it is easy to prove  that:
\begin{eqnarray*}
\hat Q\left(\begin{matrix}
\varphi_n^{+} \\
0
\end{matrix}
\right)&=&i\,\sqrt{E_n}\left(\begin{matrix}
0 \\
\varphi_n^{-}
\end{matrix}
\right)\\
\hat Q^{\dag}\left(\begin{matrix}
0 \\
\varphi_n^{-}
\end{matrix}
\right)&=&-i\,\sqrt{E_n}\left(\begin{matrix}
\varphi_n^{+} \\
0
\end{matrix}
\right)
\end{eqnarray*}
The three different kind of eigenstates (\ref{(7-1)}), (\ref{(9-1)}) and  (\ref{(9-2)}) are characterized 
by the fact that the states  in (\ref{(9-1)}) and (\ref{(9-2)})  have a well-defined ``fermion'' number $(-1)^{\mathbb F}$ while the states 
in  (\ref{(7-1)}) do not carry a well-defined one . In fact from  (\ref{(5-3)})  we get for the states (\ref{(9-1)}):
\begin{eqnarray*}
(-1)^{\mathbb F}\left(\begin{matrix}
\varphi_n^{+} \\
0
\end{matrix}\right)&=&\left[1-2\,\hat \psi\hat \psi^{\dag}\right]\left(\begin{matrix}
\varphi_n^{+} \\
0
\end{matrix}\right)\\
&=&\left[\mathbb I-2\left(\begin{matrix}
0&1\\
0&0
\end{matrix}
\right)\left(\begin{matrix}
0&0\\
1&0
\end{matrix}
\right)\right]\left(\begin{matrix}
\varphi_n^{+} \\
0
\end{matrix}\right)\\
&=&\left(\begin{matrix}
-1&0\\
0&1
\end{matrix}
\right)\left(\begin{matrix}
\varphi_n^{+} \\
0
\end{matrix}\right)=-\left(\begin{matrix}
\varphi_n^{+} \\
0
\end{matrix}\right)
\end{eqnarray*}
which means that the states $\left(\begin{matrix}
\varphi_n^{+} \\
0
\end{matrix}\right)$ have fermion number $-1$. It is easy to prove that the states  (\ref{(9-2)}) have fermion number $+1$:
\begin{equation*}
(-1)^{\mathbb F}\left(\begin{matrix}
0 \\
\varphi_n^{-}
\end{matrix}\right)=\left(\begin{matrix}
-1&0\\
0&1
\end{matrix}
\right)\left(\begin{matrix}
 0\\
\varphi_n^{-}
\end{matrix}\right)=\left(\begin{matrix}
 0\\
\varphi_n^{-}
\end{matrix}\right).
\end{equation*}
For the eigenstates (\ref{(7-1)})  we get instead:
\begin{equation*}
(-1)^{\mathbb F}\left(\begin{matrix}
\varphi_n^{+} \\
\varphi_n^{-}
\end{matrix}\right)=\left(\begin{matrix}
-\varphi_n^{+} \\
\varphi_n^{-}
\end{matrix}\right).
\end{equation*}
This implies that the states $\left(\begin{matrix}
\varphi_n^{+} \\
\varphi_n^{-}
\end{matrix}\right)$ do not carry a well-define ``fermion'' number.
The eigenstates (\ref{(9-1)}), (\ref{(9-2)}) and (\ref{(7-1)}) are also distinguished by another crucial difference: the eigenstates (\ref{(7-1)})
are entangled while (\ref{(9-1)}) and (\ref{(9-2)}) are not. It is in fact easy to see that, (\ref{(7-1)}),
\begin{equation}
\left(\begin{matrix}
\varphi_n^{+} \\
\varphi_n^{-}
\end{matrix}\right)\neq \tilde\varphi \left(\begin{matrix}
\alpha \\
\beta
\end{matrix}\right),
\label{(11-1)}
\end{equation} 
where $\tilde\varphi$ is some function of $q$ while $\alpha$ and $\beta$ are normalizable 
constant making up the spin part of the state. This proves that the states (\ref{(7-1)}) are entangled. The 
states (\ref{(9-1)})  and  (\ref{(9-2)}) are instead trivially non-entangled:
\begin{eqnarray}
\left(\begin{matrix}
\varphi_n^{+} \\
0
\end{matrix}\right)&=&\varphi_n^+\left(\begin{matrix}
1 \\
0
\end{matrix}\right)\nonumber\\
\left(\begin{matrix}
0 \\
\varphi_n^{-}
\end{matrix}\right)&=&\varphi_n^-\left(\begin{matrix}
0 \\
1
\end{matrix}\right).
\label{(12-1)}
\end{eqnarray} 
\section{Superselection Rule}
In this section we will show that actually the entangled states (\ref{(11-1)}) are not allowed because of a {\it superselection mechanism} present in the SUSYQM models. In all susy field theory models, and actually in all field theories, we are not allowed to make superpositions of bosonic states $|\phi\rangle$ with fermionic ones $|\psi\rangle$ . In fact if we make the  sum of the two states:
\begin{equation}
|\tilde\phi\rangle=|\phi\rangle+|\psi\rangle
\label{(somma)}
\end{equation} 
and then perform a rotation of $2\pi$ along the z-axis, we would not obtain  $|\tilde\phi\rangle$ but
\begin{equation}
|\tilde\phi^\prime\rangle=|\phi\rangle-|\psi\rangle.
\label{(differenza)}
\end{equation} 
This is the simplest example of superselection mechanism \cite{Wic}. In general the superselection rules are produced by the following mechanism \cite{Rom}: if there exists an operator $\hat P$ different from the identity and which commutes with all the observables of a theory, then the total Hilbert space ${\cal H}$
of the theory is naturally decomposed in the direct sum of Hilbert spaces ${\cal H}_{p_i}$ given by the eigenvarietis associated to each eigenvalue $p_{i}$ of the operator $\hat P$, called the superselection operator:
\begin{equation}
{\cal H}= {\cal H}_{p_1} \oplus\cdots \oplus{\cal H}_{p_n}. 
\label{decomposizione}
\end{equation}
The most important fact \cite{Rom} is that the allowed or "{\it physical}" states cannot be linear superpositions of states 
$|\phi_{p_i}\rangle$ belonging to different Hilbert spaces ${\cal H}_{p_i}$:
\begin{equation}
|\psi_{phys}\rangle\neq\sum \alpha_{i}|\phi_{p_i}\rangle.
\label{fisico}
\end{equation}
For more details we refer the reader to refs. \cite{Wic} and  \cite{Rom}.
\par
In the SUSYQM model a superselection operator $\hat P$ is present. It is related to the fermion number operator introduced in (\ref{(5-3)}):
\begin{equation}
\hat P=-(-1)^{\mathbb F}=\sigma_{3}
\label{uffa}
\end{equation}
Remembering the representation  (\ref{(6-1)}) of the Grassmannian variables $\hat\psi$ and $\hat\psi^{\dag}$, we have that:

\begin{equation}
{\hat P} {\hat\psi} {\hat P}^{\dag}=-{\hat\psi} ,
\label{uffa2}
\end{equation}

\begin{equation}
 {\hat P} {\hat\psi}^{\dag} {\hat P}^{\dag}=-{\hat\psi}^{\dag}. 
\end{equation}
If all the observables of the theory are even in the ${\hat\psi}$ and ${\hat\psi}^{\dag}$, then ${\hat P}$ commutes with all the observables and it then plays the role of a superselection operator. The Hilbert space get then splitted into two Hilbert spaces associated with the eigenvalues +1 and -1 of ${\hat P}$
\begin {equation}
{\cal H}= {\cal H}_{+1} \oplus{\cal H}_{-1}.
\label{uffa3}
\end{equation}
 States belonging to ${\cal H}_{+1}$ are for example the eigenstates $\left(\begin{matrix}
0 \\
\varphi_n^{-}
\end{matrix}\right)$ of $\hat H$ and those belonging to ${\cal H}_{+1}$ are for example the eigenstates  $\left(\begin{matrix}
\varphi_n^{+} \\
0
\end{matrix}\right)$ of $\hat H$. As the {\it physical} states cannot be linear superpositions of states belonging to
$ {\cal H}_{+1}$ and  ${\cal H}_{-1}$, this implies that the entangled states (\ref{(11-1)}) are not physical ones.
\par
The reader may ask why this mechanism does not act in the Jaynes-Cummings mode. The reason is that one of the observable, in particular the Hamiltonian, does not commute with $\hat P=-\sigma_{3}$. In fact the Janes-Cummings Hamiltonian contains an interaction piece of the form   \cite{ Cummings}
\begin {equation}
H_{int}=g ({\hat a}^{\dag} \sigma_{+}+{\hat a}\sigma_{-})
\label{interazione}
\end{equation}
where $g$ is a coupling, the $\sigma_{i}$ are Pauli matrices and $a,a^{\dag}$ are the annihilation and creation operators of the {\it spring} part of the action. This piece of the Hamiltonian clearly does not commute with ${\hat P}$ that is equal to $-\sigma_{3}$. So the superselection mechanism cannot work because at least one observable does not obey its rules  \cite{Rom}.
\par
If we look at (\ref{interazione})  and write it in terms of $\hat\psi$ and ${\hat\psi}^{\dag}$ we get:
\begin {equation}
H_{int}=g ( {\hat a}{\hat \psi}+{\hat a}^{\dag}{\hat\psi}^{\dag})
\label{interazione2}
\end{equation}
and from this we notice that this is an observable odd in the $\hat\psi$ and ${\hat\psi}^{\dag}$ variables.
This is something we tended to exclude in the SUSYQM model and we did that because the one above is a piece which does not conserve the "fermion" number whose conservation is one of the pillar of {\it field theory}. The usual interaction used in field theory is of Yukawa type that in SUSYQM would be:
\begin {equation}
H_{int}=g {\hat\psi}{\hat q}  {\hat\psi}^{\dag}.
\label{interazione-yu}
\end{equation}
Actually SUSYQM is not a field theory and we can relax the requirement that the "fermion" number is conserved admitting   observables with odd number of $\hat\psi$ and ${\hat\psi}^{\dag}$ like it happens in the Jaynes-Cummings model. This would  then imply that  the entangled states   (\ref{(11-1)}) are  physical ones.

\section{Conclusions}
The reader familiar with the literature on the Janes-Cummings model for sure knows that a lot of things have been explored at the interface
between this model and supersymmetry or graded superalgebras. It was first noted in \cite{rauu}  that the JC model consists of a supersymmetric harmonic oscillator plus the interaction given in eq. (\ref{interazione}) which breaks the supersummetry. Also   manners to embed the JC model in larger supersymmetric model or in superalgebras have been explored and some limited literature is contained in the papers  \cite{add,*Ras,*Niet}. Another approach has been the one of generalizing the JC model by replacing the superpotential of the harmonic oscillator with a generic superpotential
 \cite{Alex,*Shura}. 
\par In this paper of ours we do nothing of all this. We only show that in a generic SUSYQM model the physical states can be entangled, like in the JC model, but provided observables odd in the fermion numbers are allowed in the system. More work remain to be done expecially in order to give  a physical  interpretation to these observables. Some hints we are following are contained in the interesting paper of Buzano et al of ref. \cite{add,*Ras,*Niet}.
We hope to come back in the future to this topic with a longer paper of which this is just a brief report.
\begin{acknowledgments}
We wish to thank C.G.Ghirardi for his support and helpful discussions,   R.Romano for having read the manuscript,  F.Cannata, A.Gangopadhyaya, M.S. Plyushchay for guidance through the literature on SUSYQM and F.Benatti for a discussion at the begining of this project. This work has been supported by grants from INFN (GE41) and the University of Trieste (FRA 2011).
\end{acknowledgments}

\bibliography{Entanglement}

\end{document}